\documentclass[a4paper,twocolumn,showpacs,superscriptaddress,prl]{revtex4}
\usepackage{epsfig}
\usepackage{psfrag}
\usepackage{graphicx}

\begin{document}

\newcommand{\vect}[1]{{\bf #1}}
\newcommand{\comments}[1]{\hfill {\tt {eq:~#1}}} 

\title{Phase Bifurcation and Quantum Fluctuations in
Sr$_3$Ru$_2$O$_7$}

\author{A. G. Green}
\address{School of Physics and Astronomy, University of St Andrews,
North Haugh, St Andrews KY16\ 9SS, UK}

\author{S. A. Grigera}
\address{School of Physics and Astronomy, University of St Andrews,
North Haugh, St Andrews KY16\ 9SS, UK}

\author{R. A. Borzi}
\address{School of Physics and Astronomy, University of St Andrews,
North Haugh, St Andrews KY16\ 9SS, UK}

\author{A. P. Mackenzie}
\address{School of Physics and Astronomy, University of St Andrews,
North Haugh, St Andrews KY16\ 9SS, UK}

\author{R. S. Perry}
\address{School of Physics and Astronomy, University of St Andrews,
North Haugh, St Andrews KY16\ 9SS, UK}
\address{Department of Physics and International Innovation Center,
 Kyoto University, Kyoto 606-8501, Japan}

\author{B. D. Simons}
\address{Cavendish Laboratory, Madingley Road, Cambridge CB3\ OHE, UK}

\date{\today}

\begin{abstract}
The bilayer ruthenate Sr${}_3$Ru${}_2$O${}_7$ has been cited as a textbook 
example of itinerant metamagnetic quantum criticality. However, recent studies 
of the ultra-pure system have revealed striking anomalies in magnetism and 
transport in the vicinity of the quantum critical point. Drawing on fresh
experimental data, we show that the complex phase behavior reported here 
can be fully accommodated within the framework of a simple Landau theory. We 
discuss the potential physical mechanisms that underpin the phenomenology,
and assess the capacity of the ruthenate system to realize quantum 
tricritial behavior.
\end{abstract}

\maketitle 

In recent years, the field of itinerant electron metamagnetism has seen a 
resurgence of interest~\cite{Goto01,Flouquet02,
Ohmichi,Grigera01,Grigera03a,Perry04,Perry01}, 
much of it connected with quantum criticality~\cite{Sachdev}. The phenomenon 
of metamagnetism can be thought as a magnetic equivalent of a liquid-gas 
transition with the role of pressure $P$ and density being 
played by the magnetic field $H$ and magnetization. At this 
level, the $(H,T)$ phase diagram of a metamagnet translates to the well-known 
$(P,T)$ phase diagram of the liquid-gas system; a first order phase boundary 
terminating in a critical point. However, in contrast to the liquid-gas 
system, the metamagnet plays host to a crystalline lattice, the ``source'' 
of the itinerant electrons. It is this difference that is 
responsible for the interesting new phenomena that can be observed in this 
system. Firstly, the coupling of electrons to the magnetic field breaks the 
spatial symmetry of the lattice and modifies the electronic orbitals. Such 
effects play a fundamental role in shaping the effective interaction between 
the electrons. This field sensitivity can be used to tune the critical 
end-point of the first order transition to zero temperature and, thereby, 
realize a quantum critical point. Secondly, as we will show below, the 
coupling of the itinerant electron system to the lattice may, by itself, 
induce striking changes in the phase diagram.

\begin{figure}[hbt]
\centerline{\includegraphics[width=3in]{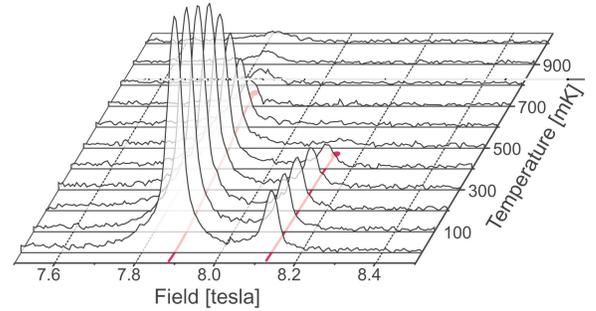}}
\caption{Measurements of the imaginary part of the a.c. susceptibility
($\chi''$) for $H\parallel c$ ($\theta=90^\circ$) as
a function of field for different temperatures. By correlating maxima
in $\chi''$ with peaks in $\chi'$ (not shown), the magnetic phase 
diagram can be inferred.
The peaks in $\chi''$ are due to dissipation associated with the crossing 
of a first order phase boundary~\protect\cite{Grigera03a}. The phase boundaries
and critical points inferred from the data at this field orientation are shown in red.} 
\label{fig:chi}
\end{figure}

The existence of metamagnetic quantum critical points (QCPs) was demonstrated 
in Sr$_3$Ru$_2$O$_7$~\cite{Grigera01} where it was shown that the field angle 
$\theta$ (measured with respect to the $ab$-plane) acts as a tuning parameter, 
allowing the construction of an $(H,T,\theta)$ phase diagram for 
metamagnetism and quantum criticality~\cite{Grigera03a}. The metamagnetic 
transition tracks a first order line in the $(H,\theta)$ plane which 
terminates at a QCP
when the energy/temperature 
scale associated with the metamagnetic transition is tuned to zero.
Recently, intriguing evidence has been reported for a new mechanism by which 
quantum critical behaviour may breakdown in samples of extremely clean 
Sr$_3$Ru$_2$O$_7$~\cite{Grigera04,Perry04}. These new 
effects are strongly dependent on purity and are seen only in the best 
crystals with residual resistivity $\rho_0 < 1 \mu\Omega {\rm cm}$. So far, 
this new behaviour has been studied mainly with the applied field oriented 
along the crystallographic $c$-axis, where the single metamagnetic QCP seen 
in lower purity samples~\cite{Grigera01,Grigera03a} is replaced by two first 
order metamagnetic transition lines at approximately $7.8$T and 
$8.1$T, each of which terminates in a finite temperature critical point with 
$T_c < 1 {\rm K}$ (see Fig.~\ref{fig:chi}). 
These lines enclose a region of anomalous transport and 
thermodynamic properties which extends up to a temperature scale of 
ca.~$1$K~\cite{Grigera04}. 
A clear correlation exists between features 
in magnetic susceptibility and magnetostriction, demonstrating a strong
magnetostructural coupling in Sr$_3$Ru$_2$O$_7$~\cite{Grigera04}.

Building on the preliminary angle-dependent resistivity data reported in 
Ref.~\cite{Grigera04}, the aim of this letter is two-fold: Firstly, we report
measurements of resistivity $\rho$ and a.c. magnetic susceptibility $\chi$ 
which confirm and extend the earlier $c$-axis data revealing an intricate 
phase diagram where the line of first order metamagnetic transitions in the 
$(H,\theta)$ plane appears to bifurcate into the two transitions observed
for $H$\,$\parallel$\,$c$. Secondly, we will show that this complex
phase behaviour can be accommodated within the framework of a Landau
phenomenology which ascribes the bifurcation to a ``symmetry-broken''
tricritical point structure. We discuss how the phenomenology, combined with 
the anomalous resistivity behaviour in the bifurcated region, 
places strong constraints on the physical mechanisms active in the 
Sr$_3$Ru$_2$O$_7$ system. Further, we address the potential experimental 
signatures of quantum tricritical behaviour.

\vskip 0.25in
\begin{figure}[hbt]
\psfrag{u}{$u$}
\psfrag{r}{$r$}
\psfrag{h}{$h$}
\centerline{\includegraphics[width=3.03in]{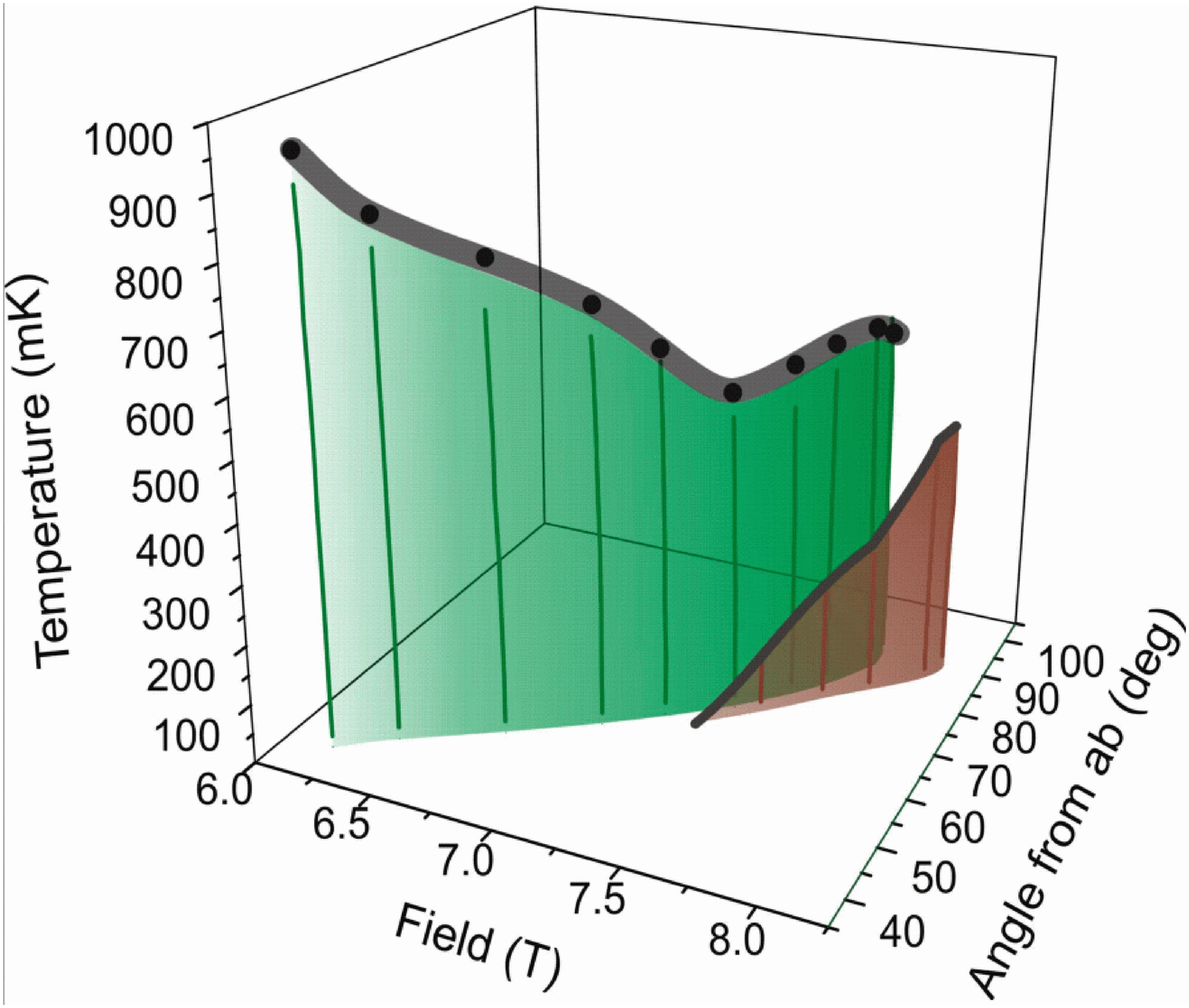}}
\vskip -3.05in
\hfill {\includegraphics[width=1.4in]{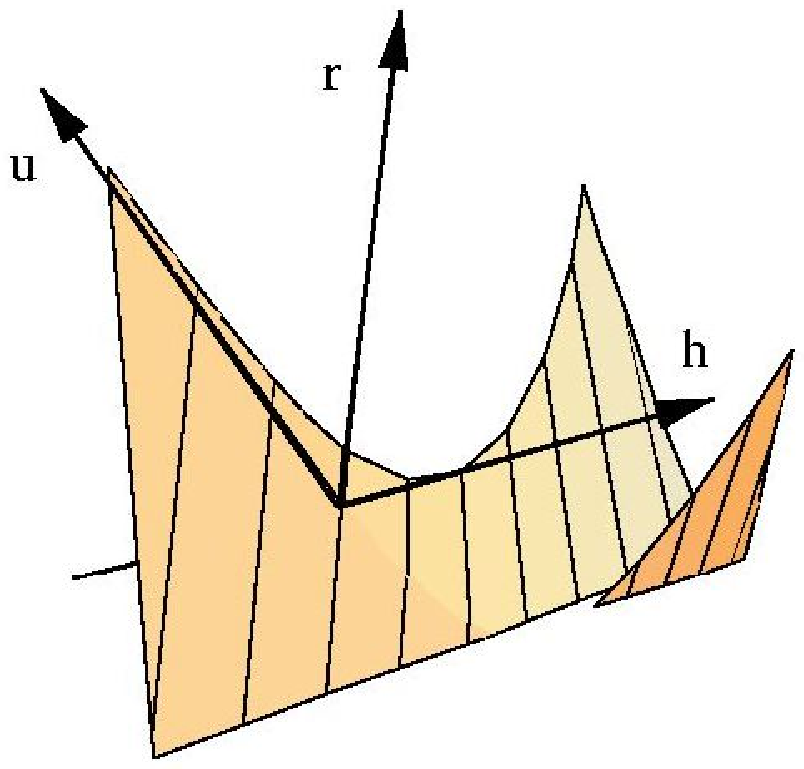}$\quad$}
\vskip 1.6in
\caption{Experimental phase diagram of ultra-pure crystals of 
Sr$_3$Ru$_2$O$_7$ as inferred from a.c. magnetic susceptibility data measured 
at $89$Hz using similar techniques to those explained in detail in 
Ref.~\cite{Grigera03a}. The planes record the loci of peaks in $\chi ''$
(cf.~Fig.~\ref{fig:chi}), 
while the thick black lines show absolute maxima in $\chi '$. The vertical 
lines and dots identify the data which have been interpolated to construct 
the figure. At each angle the field was swept through the metamagnetic region 
at fixed temperatures ranging from $50$\,mK to $1.4$K in steps of $100$\,mK.
The inset shows the phase diagram associated with the Landau 
theory~(\ref{Free_Energy1}) with the parameter $s$ and orientation chosen
to match the geometry of experiment (see main text).}
\label{fig:phaseth}
\end{figure}

\begin{figure}[hbt]
\centerline{\includegraphics[width=2.6in]{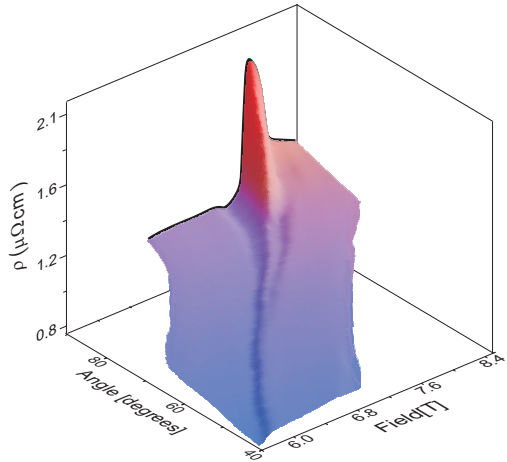}}
\caption{Resistivity data recorded at a fixed temperature $T= 100$\,mK taken
from the same range of field $H$ and angle $\theta$ as that used in 
Fig.~\ref{fig:phaseth}.}
\label{fig:resis}
\end{figure}

The new data reported here are based upon detailed
studies of the angular dependence of a.c. magnetic susceptibility ($\chi$) 
and resistivity ($\rho$). By correlating peaks in $\chi^{\prime\prime}(H,T)$ 
with absolute maxima in 
$\chi^\prime(H,T)$, the loci of first order metamagnetic transitions and
their critical end-points can be traced~\cite{Grigera03a} (see sample data in 
Fig.~\ref{fig:chi}). Fig.~\ref{fig:phaseth} shows the detailed phase 
diagram inferred from a sequence of measurements taken at different angles 
$\theta$ on high purity single crystals with $\rho_0 < 0.7 \mu\Omega 
{\rm cm}$. 
In less pure samples, the temperature of the end-point is shown to fall 
monotonically with increasing angle, and is depressed to below $100$\,mK 
at angles $\theta \gtrsim 80^\circ$\cite{Grigera03a}. Here, in the purer 
samples, one can see that the 
dependence is non-monotonic, with the critical line rising slightly in 
temperature for large $\theta$~\cite{foot1}. At the same time, a second 
surface of first order transitions emerges, with an end-point that rises 
with $\theta$. The complementary study of $\rho$,
shown in 
Fig.~\ref{fig:resis}, confirms that these first order transitions enclose
a region of anomalously high $\rho$
when $H$ is aligned very close to 
the $c$-axis \cite{Perry04,Grigera04}. Even when $\theta< 85^\circ$, where
the anomaly in the $\rho$
is weak, one can identify two distinct ridges
bifurcating from a single ridge at an angle of $\theta \approx  60^\circ$, 
a result consistent with that inferred from a $T=100\, {\rm mK}$ section 
through the magnetic phase diagram~\cite{foot2}. 

Although the magnetic phase diagram is rich, the detailed bifurcation 
structure can be accommodated within the framework of a Landau functional 
which involves the simplest generalization of the canonical theory: At the 
mean-field level, in the vicinity of a conventional metamagnetic critical
point, the Landau free energy can be expanded as $\beta F_0=hm+\frac{r}{2} 
m^2+\frac{u}{4} m^4$, where $m$ denotes the deviation of the magnetisation
density from its value $M_*$ at the critical point $h^*=r^*=0$. The
parameters $r$ and $h$ (themselves functions of $T$, $H$, and $\theta$) 
span, respectively, directions parallel and perpendicular to the line of 
first order transitions, and $u>0$. (The presence of a large external field 
in the metamagnet makes the system effectively uniaxial.) However, if the 
sign of the interaction $u$ is reversed, one is compelled to consider the 
generalization,
\begin{equation}
\beta F[m]=hm+\frac{r}{2} m^2+\frac{s}{3} m^3 + \frac{u}{4} 
m^4 + \frac{1}{6} m^6,
\label{Free_Energy1}
\end{equation}
where the presence of the cubic term in $m$ reflects the fact that, in the 
metamagnetic system, only the quintic term may be removed by rescaling. 

\begin{figure}[hbt]
\psfrag{(a)}{$(a)$}
\psfrag{(b)}{$(b)$}
\psfrag{u}{$u$}
\psfrag{u1}{$u$}
\psfrag{r}{$r$}
\psfrag{r1}{$r$}
\psfrag{h}{$h$}
\psfrag{h1}{$h$}
\psfrag{T}{}
\psfrag{P}{$P$}
\centerline{\includegraphics[width=3.0in]{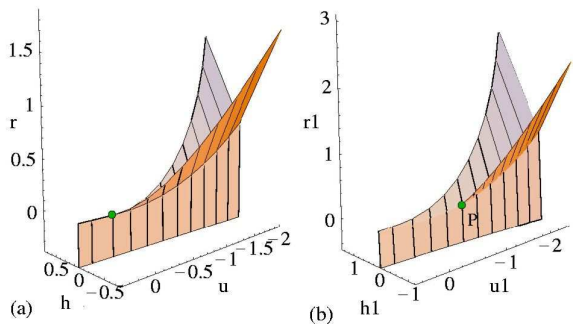}}
\caption{\small Phase diagram of the Landau theory~(\ref{Free_Energy1}) 
with (a) $s=0$ and (b) $s=0.2$. The planes of first order transitions 
are terminated by lines of critical points. Note that, for $s=0$, the 
point of bifurcation occurs at the tricritical point: $h_*=r_*=u_*=0$ 
while, for $s\ne 0$, the point of bifurcation becomes dislocated. The zero 
temperature plane of the physical system forms a plane that bisects the 
critical lines near the region of bifurcation, with the second critical 
line emerging at some negative value of temperature (see 
Fig. \ref{fig:phaseth} inset).}
\label{fig:phase}
\end{figure}

To understand how the phase behaviour is recovered from (\ref{Free_Energy1}),
it is instructive to consider first a ``symmetric'' theory with $s=0$. In this 
case, a change in the sign of $u$ leads to tricritical 
phenomena~\cite{Griffiths70,note}. As shown in Fig.~\ref{fig:phase}a, the 
phase diagram is characterised by a bifurcation of the critical line at the 
tricritical point: $h^*=u^*=r^*=0$. For $u>0$, the critical line bounds a 
plane of first order transitions while, for $u<0$, two critical lines 
bound first order planes which coalesce into a single plane along a line of 
degeneracy. The trajectories of the bifurcated critical lines for $u<0$ are 
given by $h^*(u)=\pm 6u^2(3|u|/10)^{1/2}/25$, $r^*(u)=9u^2/20$ while the line 
of degeneracy follows a trajectory $h_{\rm deg}=0$, $r_{\rm deg}(u)=3 u^2/16$. 
Restoring the cubic contribution, the point of bifurcation becomes 
`dislocated' such that the second line of critical points emerges from the 
plane of first order transitions at a point P: $u_{\rm P}(s)=-(4 s)^{2/3}$, 
$h_{\rm P}=su_{\rm P}/4$, $r_{\rm deg}=3u_{\rm P}^2/16$
while, away from the region of bifurcation, an expansion in $s$ shows the 
critical lines to asymptote to the trajectories $h^*(u,s)\simeq h^*(u)+2us/5$, 
$r^*(u,s)\simeq r^*(u)\pm s\sqrt{-6u/5}$ (see Fig.~\ref{fig:phase}b). The 
corresponding line of degeneracy follows the trajectory $h_{\rm deg}=
su/4$, $r_{\rm deg}=3u^2/16$ .

When compared with the bifurcation structure of the measured phase diagram,
the correspondence with the Landau theory~(\ref{Free_Energy1}) is striking: 
The bifurcation in the experimental system is consistent with the second 
line of critical points emanating from a point P located at some small 
negative temperature and rising through the $T=0$ plane at an angle of 
ca.~$80^\circ$ (see Fig.~\ref{fig:phaseth}, inset). Moreover, the primary 
line of critical points shows an upturn which is also predicted by the 
Landau theory. 

To go beyond the level of phenomenology, it is necessary to identify physical 
mechanisms which may be responsible for reversing the sign of $u$. Although 
one cannot rule out idiosyncrasies of the electron band structure 
influencing the coefficients of the Landau expansion, such effects taken 
alone would be unlikely to explain the extreme sensitivity to disorder and 
the anomalous resistivity dependence observed in the bifurcated region. 
However, the Landau coefficients may be adjusted indirectly by coupling the 
magnetization density to some auxiliary 
field~\cite{Rice54,Garland66,Larkin1969},
\begin{equation}
\beta F[m,\psi]=\beta F_0+\gamma (M_*+m)^2\psi+\beta F_\psi[\psi].
\label{Larkin_Pikin}
\end{equation}
Crucially, when integrated out, fluctuations of the field $\psi$ impart 
a \emph{negative} contribution to the quartic interaction of $m$; $u\mapsto 
u-4\gamma^2\langle \psi^2\rangle$. What physical mechanisms could give rise
to such a coupling?

In fact, there are relatively few choices available whose symmetry allows the 
simple coupling to magnetization given by~(\ref{Larkin_Pikin}). Correlations 
of the magnetostriction data with magnetic susceptibility reported in the
$\theta=90^\circ$ system~\cite{Grigera04} 
suggest an association of the field $\psi$ with the 
lattice strain. (Indeed, this was the coupling originally considered in 
Refs.~\cite{Rice54,Garland66,Larkin1969}.) However, a mechanism driven solely 
by harmonic fluctuations of the lattice sits uncomfortably with the observed 
disorder dependence of the bifurcation; the bifurcation appears only in the 
pure system and is quenched by tiny amounts of disorder, while the effect of 
disorder upon lattice fluctuations is likely to be weak. 
This difficulty may be resolved by drawing upon the physical origin of 
metamagnetism: By effecting an increased magnetic polarization, the Fermi 
level can be positioned in a region of high density of states (DoS). By 
combining metamagnetism with a weak structural transition or, potentially, 
an interaction-driven Fermi surface distortion, the system may take 
energetic advantage of singular features in the local (in $k$-space)
DoS~\cite{Grigera04} (similar to a Jahn-Teller or Peierls distortion in an 
insulator). Approaching the critical point of the undistorted system, a 
lattice or Fermi surface distortion could split the peak in the DoS and 
thereby advance the transition. 
Elastic scattering of electrons from impurities would smear out the features 
in the DoS that provide the energetic drive and so quench the bifurcation. 
Note that, due to the restrictions of symmetry, the Landau theory will take 
precisely the same form whether one chooses to identify $\psi$ with the 
lattice strain or with the size of a Fermi surface distortion; indeed, structural distortions
are inevitably accompanied by Fermi  surface distortions.

As well as capturing both the observed features in magnetostriction and the 
quenching of the bifurcation by 
disorder, such a mechanism 
affords a natural explanation for the resistance anomaly: A degeneracy of 
stable lattice configurations or Fermi surface distortions would be 
accompanied by the nucleation of ordered domains~\cite{Grigera04,foot-landau}. 
Once the (diverging) magnetic correlation length exceeds the domain size, the 
resistivity will become controlled by the scattering from the latter.
Further, one may expect that the tilting of the magnetic field away from the 
$c$-axis would lift the degeneracy, destroying the domain structure and, with
it, the peak in resistivity. 

To close the discussion, let us turn to a consideration of further observable 
consequences of the Landau theory~(\ref{Free_Energy1}). In the symmetric 
($s=0$) theory, the development of a tricritical point is accompanied by a 
substantial softening of classical and quantum fluctuations. In the present 
case, where $s\ne 0$, when the temperature exceeds the `energy scales' of 
the bifurcation region~\cite{foot_energy_scales} (ca.~$1$\,K), the 
fluctuations will remain characteristic of a quantum tricritical point 
(see below). As the temperature is reduced, the system will pass through 
two further regimes of behaviour: At the lowest temperatures, all magnetic 
fluctuations will be gapped. Between these high and low temperature 
extremes, the system will pass through a crossover regime where the
behaviour will be determined by the proximity to the finite temperature 
critical points.

To address the behavior at higher temperatures, where fluctuations are 
controlled by a quantum tricritical point, one can employ an extended 
Hertz-Millis action~\cite{Hertz_Millis},
\begin{eqnarray*}
{\cal S}&=&\frac{1}{2}\int\frac{d {\bf q}}{(2 \pi)^3} \frac{d 
\omega}{2\pi}\left(r_0 + {\bf q}^2+ \frac{|\omega|}
{\Gamma_{\bf q}}\right)|m({\bf q},\omega)|^2\nonumber\\
&&\qquad +\int d{\bf r}\, d\tau\left(u_0 m({\bf r},\tau)^4+
v_0 m({\bf r},\tau)^6\right),
\end{eqnarray*}
where $u_0<0$, $v_0>0$ and $\Gamma_{\bf q}=v|{\bf q}|$. To leading order 
in the bare parameters $r_0$, $u_0$, and $v_0$, the influence of fluctuations 
can be incorporated by following a self-consistent renormalization 
procedure~\cite{Moriya1985} from which one obtains
\begin{eqnarray*}
\left\{\begin{array}{l}
r(T) = r_0 + 12 u_0 \langle m^2 \rangle + 90 v_0 \langle m^2 \rangle^2\cr
u(T) = u_0 + 15 v_0 \langle m^2 \rangle \cr
v(T) = v_0.\cr
\end{array}\right.
\end{eqnarray*}
Here the averages $\langle\cdots\rangle$ are calculated self-consistently 
with the renormalized action. To leading order,
one need retain only $r(T)$, and perform the calculation with the 
corresponding quadratic action. Subtracting zero-point fluctuations, 
one obtains $r(T)=r(0)+12 u(0)(\langle m^2 \rangle-\langle m^2 \rangle_{T=0})
+ 90 v(0)(\langle m^2 \rangle-\langle m^2 \rangle_{T=0})^2$,
%
%
where
\begin{eqnarray*}
\langle m^2 \rangle-\langle m^2 \rangle_{T=0}
=\left\{\begin{array}{ll}
\frac{T^{4/3} \Gamma^{-1/3}}{\pi^2 \sqrt{3}} & T \gg r(0)\cr
\frac{T^2}{\pi^3 \Gamma r(0)} & T \ll r(0)\cr
\end{array}\right.,
\label{flucts}
\end{eqnarray*}
denotes the thermal contribution to the critical fluctuations. At a 
conventional quantum critical point, $r(0)=0$ and $u(0)>0$, leading to the 
characteristic temperature dependence $r(T) \propto 
T^{4/3}$~\cite{Hertz_Millis}. By contrast, at the quantum tricritical point, 
$r(0)=u(0)=0$, leading to an enhanced temperature dependence 
$r(T)\propto T^{8/3}$ reflecting the shallow potential for fluctuations. 
Translated to the electron self-energy, the magnon fluctuations contribute 
a factor ${\rm Im}\, \Sigma^R({\bf k},0) \sim 
T^3/
r(T)$ from which 
one infers a resistivity of $\rho \propto T^{5/3}$ for a conventional quantum 
critical point and $\rho \propto T^{1/3}$ for the quantum tricritical point. 

To conclude, we have shown that the bifurcation structure observed in the 
magnetic susceptibility of Sr${}_3$Ru${}_2$O${}_7$ is consistent with a 
Landau phenomenology reflecting a `dislocated' tricritical point structure. 
Further, we have argued that, by coupling lattice fluctuations to a Fermi 
surface instability, the Landau phenomenology provides a natural explanation 
of the resistance anomaly. It is interesting to note that the bifurcation 
mechanism described here is quite generic and may provide an opportunity to 
realize quantum \emph{tricritical} behavior both in Sr${}_3$Ru${}_2$O${}_7$ 
and potentially more widely. Indeed, even within the ruthenate
system, there is growing evidence that the neighboring metamagnetic 
transitions revealed in the pure system are also accompanied by bifurcation 
structures. The distortion of the Fermi surface through lattice instabilities 
or strong interactions may provide a general mechanism for clean materials 
to mask a magnetic quantum critical point.  

{\sc Acknowledgments:} We are grateful to Nigel Cooper, Peter Littlewood and 
Gil Lonzarich and Yoshiteru Maeno for valuable discussions.



\begin{thebibliography}{}
\bibitem{foot2} At first sight there may appear to be a discrepancy between 
Figs.~\ref{fig:phaseth} and \ref{fig:resis}. The resistivity data of 
Fig.~\ref{fig:resis} show a clear bifurcation at $\theta = 60^{\circ}$ and 
$T = 100\,$mK whilst, within the angular resolution of $5^{\circ}$, it is 
not possible to say whether or not the first order planes of 
Fig.~\ref{fig:phaseth} intersect at $100\,$mK. We stress that, even if the 
first order planes of Fig.~\ref{fig:phaseth} do not intersect at $100\,$mK, 
the figures are consistent. Unlike $\chi''$, $\rho$ can show some signal 
when a crossover rather than a first-order transition is taking place, so 
features in $\rho$ can extend beyond the first-order planes of 
Fig.~\ref{fig:phaseth}.
\end{thebibliography}
\end{document}